\documentclass[12pt]{article}
\usepackage{latexsym,graphicx,multirow}
\usepackage{float}
\usepackage{amssymb}
\usepackage{amsmath}
\usepackage{amscd}
\usepackage{amsthm}
\usepackage[left=2cm,top=2.5cm,right=2.5cm,bottom=1.5cm]{geometry}
\usepackage{hyperref}
\usepackage{epstopdf}
\usepackage{cite}
\usepackage{caption}
\usepackage{subcaption}
\usepackage{color}
\begin{document}
	
\begin{center}
\large{\bf{Observation constraints on scalar field cosmological model in anisotropic universe}} \\
		\vspace{10mm}
\normalsize{Vinod Kumar Bhardwaj$^1$, Anil Kumar Yadav$^2$  }\\
		\vspace{5mm}
\normalsize{$^{1}$Department of Mathematics, GLA University, Mathura-281 406, Uttar Pradesh, India}\\
		\vspace{2mm}
\normalsize{$^{3}$Department of Physics, United College of Engineering and Research, Greater Noida - 201310, India.}\\
		\vspace{2mm}
$^1$E-mail:dr.vinodbhardwaj@gmail.com\\
		\vspace{2mm}
$^2$E-mail:abanilyadav@yahoo.co.in\\
		\vspace{10mm}
		
		%\date{}
		%\maketitle
\end{center}
	
\begin{abstract}
	In this study, we have explored a scalar field cosmological model in the axially symmetric Bianchi type-I universe. In this study, our aim is to constrain the scalar field dark energy model in an anisotropic background. For this purpose, the explicit solution of the developed field equations for the model is determined and analyzed. Constraints on the cosmological model parameters are established utilizing Markov Chain Monte Carlo (MCMC) analysis and using the latest observational data sets of OHD, BAO, and Pantheon. For the combined dataset (OHD, BAO, and Pantheon), the best-fit values of Hubble and density parameters are estimated as $ H_{0} = 71.54\pm 0.28$, $\Omega_{m0}=0.2622\pm0.0021$  $\Omega_{\phi0}  = 0.7331\pm0.0046$, and $\Omega_{\sigma 0} = 0.000162\pm0.000063$. The model shows a flipping nature and redshift transition occurs at $z_{t} = 0.6964^{+0.0136}_{-0.0006}$, and the present value of decelerated parameter is computed to be $q_{0} = -0.6964\pm0.028$ for the combined dataset. We have explored characteristics like the universe's age, particle horizon, deceleration parameter, and jerk parameter.  The dynamical properties such as energy density $\rho_{\phi}$, scalar field pressure $p_{\phi}$, and equation of state parameter $\omega_{\phi}$ are analyzed and presented.  We have also described the behavior of the scalar potential $V(\phi)$ and scalar field.  Furthermore, the authors also described the behavior of energy conditions in scalar-tensor cosmology. The scenario of the present accelerated expansion of the universe is described by the contribution of the scalar field.
\end{abstract}


%%%%%%%%%%%%%%%%%%%%%%%%%%%%%%%%%%%%%%%%%%%%%%% Section 1 %%%%%%%%%%%%%%%%%%%%%%%%%%%%%%%%%%%%%%%%%%%%%%%%%%%%%%%%%%%%
\section{Introduction}
Einstein established the general relativity (GR) theory in the early 20th century to depict the connection of gravity with space and time. He presented a direct link of matter and radiation with space-time geometry. Connecting the energy and momentum, inherent properties of matter and radiation with space-time curvature Einstein formulated field equation is read as $R_{ij}-\frac{1}{2}R g_{ij} =- 8\pi G T_{ij}$ \cite{ref1}. The cosmic models explaining the present scenario of the expanding universe are based on the general relativity theory. The current accelerated era of expanding universe is also validated by several experimental studies \cite{ref2,ref3,ref4,ref5}. The dark energy (DE) is expected as main contributor in universe’s expansion while dark matter is supposed to be primary element for development of large-scale structures (LSS)\cite{ref6,ref7,ref8}. The dark energy with huge amount of repulsive (negative) pressure is an enigmatic form of energy which is responsible for universe’s expansion. These perceptions and confirmations about the cosmos motivate the theoretical researchers to develop the models of the universe in distinct frameworks. Due to its repulsive nature, the cosmological constant has been assumed to be an appropriate substitute for DE \cite{ref9}. In 2012, Bamba et al. \cite{Bamba/2012} reviewed dark energy cosmology by analyzing different theoretical models and cosmography tests, and finally, they showed that the degeneration among parameters can be removed by accurate data analysis of large data samples. Furthermore, we note that in order to reveal the cosmological constant problems, the cosmological models with variable equations of state parameters for dark energy \cite{Akarsu/2010, Kumar/2011, Kumar/2011a, Yadav/2011} and the holographic dark energy (HDE) models \cite{Nojiri/2004, Nojiri/2005, Nojiri/2006, Nojiri/2005a} have been investigated. In 2011, Nojiri and Odintsov \cite{Nojiri/2011} explained the cosmic history of the universe from its early phase to the present epoch. Recently, Nojiri et al. \cite{Nojiri/2017} have investigated a cosmological model that successfully described the late-time dynamics of the universe. \\

Many alternate theories without cosmological constant (CC) have been suggested in literature to describe the current accelerated expansion of the universe. Some of the modified theories that have been proposed Einstein-Gauss-Bonnet theory, $f(T)$, $f(R)$, $f(R,T)$,$f(R,G)$, and $f(Q)$ gravities \cite{ref10,ref11,ref12,ref13,ref14}. All these theories make different predictions for the behavior of the universe and the properties of DE. Some of these theories also have additional parameters that can be tuned to fit observational data. Although the cosmological constant nicely fits the experimental findings and is supported by a wide range of experiments but false to describe the inflation era of the universe. The Weyl, Lyra, Brans-Dicke, and others theories are a few examples of modified theories for describing late-time cosmic expansion without using cosmological term, see ref.\cite{ref15,ref16,ref17,ref18}. The modified gravity, which eliminates the need for dark energy and which seems to be stable, is investigated \cite{Nojiri/2011, Nojiri/2017, Nojiri/2003prd}. Nojiri and Odintsov \cite{Nojiri/2011} explained the cosmic history of the universe from its early phase to the present epoch. Also, Nojiri et al. \cite{Nojiri/2017} have investigated a cosmological model that successfully described the late-time dynamics of the universe. One of the major research that describes the modified theory of gravity with negative and positive powers of the curvature.
\\

In addition to modified theories, some cosmological models based on the scalar field theory are also formulated theoretically to describe the inflation era as well as the universe’s current expansion phase \cite{ref19,ref20}. These models assume that the dark energy component is a scalar field ($\phi$). The negative pressure is generated by a scalar field along with a decreasing potential ($V(\phi)$). Scalar field cosmologies have been proposed in several studies to explain the dynamics of the universe\cite{ref19,ref20,ref21,ref22}. According to references \cite{ref21,ref22}, the quintessence model is a more palatable scalar field cosmic model that accurately avoids the classic problems of fine-tune and cosmic coincidence, and describes the current cosmic reality. The notion of tracking was first put forth by Johri \cite{ref23}, who suggested a definite course for the universe’s evolution due to the tracker’s potential. The observational estimates provided strong support for the idea. In literary works, various quintessence models have been suggested. These ideas include the possibility of a scalar field development driven by a non-canonical kinetic term \cite{ref24} and the non-minimal relationship between quintessence and dark matter \cite{ref25,ref26}. Significant uses of time dependent equation of state (EoS) parameter in scalar tensor theory are mentioned in \cite{ref27,ref28}. Many fundamental theories also acknowledge the scalar field's presence in astrophysical studies. Recently, many cosmic models have been proposed in different framework of scalar field theory \cite{ref27,ref28,ref29,ref30,ref30a}. The dark energy model like Chaplygin gas with unique EoS parameter has been studied by Kamenshchik et al. \cite{ref31}.\\

Among the cosmological events, expansion of universe is predicted and well evidenced from latest observations \cite{ref2,ref32}. To describe these phenomenons occurred in cosmological scenarios, different models have been proposed through modifying the field equations. Initially the modeling is done by considering universe as isotropic and homogeneous in nature for example in case of models based FRW theory \cite{ref26,ref27,ref28,ref29,ref30}. Later on with the analysis of CMB radiations data \cite{ref5}, disputes aroused about isotropic nature of universe, especially at its early stage. Observations made by WAMP also favoured anisotropic universe\cite{ref5,ref33,ref34,ref34a,ref34b}. Considering the CMB radiation predictions along with WAMP confirmations, Bianchi's formulation of geometrical space seems suitable to describe anisotropy of universe. In this regard various models based on anisotropic Bianchi formulations have been proposed to discuss cosmological events \cite{ref35,ref36,ref37,ref38,ref39}. Recently  Nojiri et al. \cite{ Nojiri/2022NPB} have formalized anisotropic evolution in the context of modified gravity, focusing on pre-inflationary and near the vicinity of the inflationary epochs. In this study, a scalar field model has been explored in axially symmetric Bianchi I universe. To extract the best value model parameters and validation, we used data from Hubble observations, BAO and Pantheon compilation of SN Ia. The manuscript is arranged as follows: The model and its solutions are discussed in section 2. We have discussed statistical technique and estimate the best fit values of parameters in section 3. Model's features and dynamical properties are conferred in section 4. In conclusion section, we have proposed a brief summary of model.
%%%%%%%%%%%%%%%%%%%%%%%%%%%%%%%%%%%%%%%%    	
\section{Basic Formalism of the model}
The equation of action for the scalar field is consider as

\begin{equation}\label{1}
S =  \int {d^4 x \sqrt{-g} \bigg[ \frac{c^4}{16 \pi G} R +\biggl\{\frac{1}{2} g^{ij} \phi_{i} \phi_{j} - V(\phi) \biggl\} +\mathcal{L} \bigg]}
\end{equation}
where, $R$, $G$ and all the other terms have standard meaning.
With the assumption $8 \pi G = c = 1$, the expression of action for matter and scalar field is read as   
\begin{equation}\label{2}
S_{\phi} =  \int {d^4 x \sqrt{-g} \biggl[\frac{1}{2} g^{ij} \phi_{i} \phi_{j} - V(\phi) \biggl] }
\end{equation}
\begin{equation}\label{3}
S_{m} =  \int {\mathcal{L} \  \sqrt{-g} \ d^4 x}
\end{equation}
The Einstein's equation can be re-write as 
\begin{equation}\label{4}
R_{ij}-\frac{1}{2} R g_{ij} = - T_{ij} -\phi_{i} \phi_{j} + g_{ij} \left(\frac{1}{2} \phi^{k}\phi_{k}-V(\phi)\right)
\end{equation}
Note, the tensors of energy momentum for the scalar field and perfect fluid are $ T^{ij}_{\phi} = (\rho_{\phi}+p_{\phi}) u_{i} u_{j} - p_{\phi} g_{ij}$, and $ T^{ij}_{m} = (\rho_{m}+p_{m}) u_{i} u_{j} - p_{m} g_{ij}$ respectively.\\

For the purpose of modelling, we are adopting the metric of locally rotationally symmetric (LRS) Bianchi type I  having the following form:

\begin{equation}\label{5}
ds^2=A^2 (dx)^2+ B^2 \left( dy^2+ dz^2 \right)-dt^2
\end{equation}
Here all the terms have standard meaning with A and B as time dependent variables. Considering above mentioned assumptions the field equations for Bianchi type I metric will adopt the following from \cite{ref38,ref40}:  
\begin{equation}\label{6}
2 \frac{\ddot{B}}{B}+ \frac{\dot{B}^2}{B^2} = - p_{\phi}
\end{equation}
\begin{equation}\label{7}
\frac{\ddot{A}}{A}+\frac{\ddot{B}}{B}+ \frac{\dot{A} \dot{B} }{A B}  = - p_{\phi}
\end{equation}
\begin{equation}\label{8}
2\frac{\dot{A} \dot{B}}{AB}+\frac{\dot{B}^2}{B^2} = \rho_{\phi}+\rho_{m}
\end{equation}
Here in equations (6)-(8) the symbols $p$ and $\rho$ represent pressure and energy density  of the universe respectively. Whereas, the subscripts $m$ and $\phi$ attached with $p$ and $\rho$ describe the contribution made by perfect fluid and scalar field respectively. 

Since the scalar field $\phi$ is function of time, so it can be assumed that $p_{\phi}$ and $\rho_{\phi}$ respectively are the pressure and energy density for perfect fluid. 
On taking scalar field $\phi$ as the single self-interacting dark energy source with potential $V(\phi)$, then in the context of LRS Bianchi-I cosmology, the
energy density $\rho_\phi$ and pressure $p_\phi$ in terms of $\phi$ are read as
\begin{equation}\label{9}
\rho_{\phi} = \frac{\dot{\phi}^2}{2} +V(\phi)
\end{equation}
\begin{equation}\label{10}
p_{\phi} = \frac{\dot{\phi}^2}{2} - V(\phi)
\end{equation}
Here, $V(\phi)$ and $\frac{\dot{\phi}^2}{2}$ respectively, denote the potential and kinetic energies. For the scalar field, the EoS parameter is given by $p_{\phi} = \omega_{\phi} \rho_{\phi} $ and potential can be read as $ V(\phi) = \frac{1-\omega_{\phi}}{2(1+\omega_{\phi})} \dot{\phi}^{2}$.\\

\noindent For the considered cosmos system, volume will be $AB^2$ with scale factor $a = (AB^2)^{1/3}$, and Hubble parameter $ H = \frac{1}{3} \left(\frac{\dot{A}}{A}+2\frac{\dot{B}}{B}\right)= \frac{\dot{a}}{a} $.
Taking the dust filled universe ($p_m = 0 $) into account, the energy conservation equation Eq. (8) will reduces to  
\begin{equation}\label{11}
\frac{d}{dt}(\rho_{m}+\rho_{\phi}) + 3 (\rho_{m}+\rho_{\phi}+p_{\phi}) H = 0
\end{equation}
Further assuming the no energy exchange among the contributing sectors of cosmos \cite{ref41,ref42}, we can write $\frac{d}{dt}(\rho_{m}) + 3 H \rho_{m} = 0$ and $\frac{d}{dt}(\rho_{\phi}) + 3 H (\rho_{\phi}+p_{\phi}) = 0$.\\
Simplifying  Eq.(6), (7) and (8), with the relation $\frac{\dot{A}}{A}+2\frac{\dot{B}}{B} =3\frac{\dot{a}}{a}$, the expression of Hubble parameter for the proposed model can be developed in the following form: 
\begin{equation}\label{12}
H^2 =\frac{1}{3}\left(\rho_{m}+ \rho_{\phi}+ \frac{1}{3} \frac{c^{2}}{a^6}\right)
\end{equation}
Here in Eq. (10), $c$ is the constant of integration evolved during mathematical simplification. Equation (10) represents Hubble parameter in time domain as a function of scaling factor $a$. 
In terms of redshift $z$, the scale factor $ a $ can be realized as $\frac{a_{0}}{a} =1+z $. In the derived model present value of scale $a_{0}$ is taken as 1 \cite{ref37,ref42}. The value of  For  Following the dependencies of various cosmos sector energy density on $z$  as discussed in \cite{ref37,ref42,ref43} with anisotropy energy density $\frac{1}{3} \frac{c_{1}^{2}}{a^6}  =\rho_{\sigma} =\rho_{\sigma0} (1+z)^{6}$, the Hubble parameter can be transformed to
\begin{equation}\label{13}
H^2 =\frac{1}{3}\left(\rho_{m0} (1+z)^3 +\rho_{\phi0} (1+z)^{3 (1+\omega_{\phi})}+ \rho_{\sigma0} (1+z)^{6}\right)
\end{equation}
Equation (13) can be further simplified to a normalized form utilizing critical density of universe $ \rho_{c} = \frac{3 H^{2}}{8 \pi G} $ and approximating $\ 8 \pi G $ to $ 1 $ as follow:
\begin{equation}\label{14}
H^2 = H^{2}_{0}\left[ (1+z)^{3}\Omega_{m0}+\Omega_{\phi0} (1+z)^{3(1+\omega_{\phi})}+ (1+z)^{6}\Omega_{\sigma0}\right]
\end{equation}

Here the term $\Omega_{i}$ represents $\frac{\rho_{i}}{\rho_{c}}$ and the subscripted $\Omega_{i0}$ its value at $z=0$. Thus for $z=0$ the Eq. (12) reduces to \cite{ref37,ref42,ref43} -
\begin{equation}\label{15}
1=\Omega_{m0} +\Omega_{\phi0}+\Omega_{\sigma0} 
\end{equation}

To get explicit solution for the proposed model, an extra assumption or parameterization must be taken into account. In this regard, we have followed the parametrization proposed by Chevallier and Polarski \cite{ref44}, and Linder \cite{ref45}

\begin{equation}\label{16}
\omega_{\phi} =\omega_{0} +\omega_{1} \left(\frac{z}{1+z}\right) 
\end{equation}
%The continuity equation for scalar field can be rewrite as
The scalar field continuity relation can be rewrite as
\begin{equation}\label{17}
3 \,\dot{\phi}^{2} \frac{\dot{a}}{a}  +\frac{d}{dt} \bigg[V(\phi)+\frac{1}{2}\dot{\phi}^{2}\bigg] = 0
\end{equation}
From Eqs. (16) and (17), we obtain
\begin{equation}\label{18}
\dot{\phi}^{2} = \dot{\phi_0}^{2} \left(\frac{(1+\omega_{\phi})(1+z)+\omega_{1} z}{(1+\omega_{0})(1+z)}\right) (z+1)^{3 \left(\omega_{0}+\omega_{1}+1\right)} e^{-\frac{3 \left(\omega _1 z\right)}{z+1}}
\end{equation}
where $\dot{\phi_0}^{2}$ represents the value of $\dot{\phi}^{2}$ at $z=0$.\\

For scalar field, following are the expressions of density and pressure respectively
\begin{eqnarray}\label{19}
\rho_{\phi} & = & \frac{1}{2} \dot{\phi}^{2} + V(\phi) = \rho_{\phi 0}  (z+1)^{3 \left(\omega_{0}+\omega_{1}+1\right)} e^{-\frac{3 \left(\omega _1 z\right)}{z+1}}
\end{eqnarray}
\begin{eqnarray}\label{20}
p_{\phi} & = & \omega_{\phi} \rho_{\phi}  = \frac{\rho_{\phi 0}\big((1+\omega_{0})(1+z)+\omega_{1}z\big)}{(1+z)}  (z+1)^{3 \left(\omega_{0}+\omega_{1}+1\right)} e^{-\frac{3 \left(\omega _1 z\right)}{z+1}}
\end{eqnarray}

From (14) and (16), the expression for the Hubble parameter is re-framed as
\begin{equation}\label{21}
H^{2} = H_0^2 \bigg[\Omega_{m 0} (z+1)^3 +\Omega_{\phi 0} (z+1)^{3 \left(\omega_{0}+\omega_{1}+1\right)} e^{-\frac{3 \left(\omega _1 z\right)}{z+1}}+\Omega_{\sigma 0} (z+1)^6 \bigg]
\end{equation}
Deceleration parameter is defined as $q= -\frac{\ddot a}{a H^2} = -1 + \frac{(1+z)}{H(z)} \frac{d H(z)}{dz} $. Thus, for the proposed model, the expression of DP is re-framed as:
\begin{eqnarray}\label{22}
q =\frac{(z+1) e^{\frac{3 \omega _1 z}{z+1}} \left(\Omega_{m0}+4 (z+1)^3 \Omega_{\sigma 0}\right)+\left(3 \omega_1 z+3 \omega_0 (z+1)+z+1\right)\xi}{2 (z+1) \left(e^{\frac{3 \omega_1 z}{z+1}} \left(\Omega _{m0}+(z+1)^3 \Omega _{\sigma 0}\right)+ \xi\right)}
\end{eqnarray}
where $\xi = (z+1)^{3 \left(\omega _0+\omega _1\right)} \Omega_{\phi 0}$.\\

The Hubble parameter is a crucial quantity in the explanation of the universe's expansion and is helpful in determining the universe's age. The deceleration parameter, on the other hand, specifies the phase change (acceleration or deceleration) that occurs during the evolution of the universe.\\
%%%%%%%%%%%%%%%%%%%%%%%%%%%%%%%%%%%%%%%%%%%%%%%%%%%%%%%%%%%%%%%%%%%%%%%%%%%%%%%%
\section{Extraction of parameters of the model} 
The experimental data and statistical methods can be used to evaluate the free parameters collected in the preceding section. Let $O_e$ be the experimental observed value and $O_t$ be the corresponding theoretically estimated value for any observable $O$. Here, $O_t$ depends on the proposed model's physical variable parameters. By comparing the two values ($O_e$ and $O_t$) with statistical methods, the model's parameters can be estimated. In order to get the best estimated values, we employed the $\chi^2$ estimator. If $\sigma$ denotes the is the standard error in the experiential values. The expression of $\chi^2$ estimator is given as
\begin{equation}\label{23}
\chi^{2}=\sum_{}^{} {\frac{\left(O_{t}-O_{e}\right)^2}{\sigma^2}}
\end{equation}
%Here, the Markov Chain Monte Carlo (MCMC) method based on Metropolis-Hastings algorithm has been implemented to determine the best fit values of parameters \cite{ref46}. 
Here, the most plausible values of parameters are determined using Markov Chain Monte Carlo (MCMC) method \cite{ref46}. By minimizing the value of $\chi^2$ estimator for any observational data \cite{ref47}, the best-fit values can be determined.
In the present study, we have employed Observation Hubble Data (OHD), Pantheon and Baryon Acoustic Oscillation (BAO) data. We have used 57 OHD points with $z$ ranging in $[0.07,2.36]$ \cite{ref48}. In case of Pantheon, 1048 supernovae apparent magnitudes  are taken into account with $z$ limited to set $[0.01,2.6]$ \cite{ref49}. The BAO data utilized for fitting is taken from the references \cite{ref50,ref51,ref52,ref53}. Now we are ready to extract free parameters, $H_0$, $\Omega_{m0}$, $\Omega_{de0}$, $\omega_{de}$, $\Omega_{\phi0}$ and $\Omega_{\sigma0}$ of the proposed model.\\
For the approximation of free parameters we have considered two cases: (i) Parameters approximated by minimizing the $\chi^2$ of individual data sets and (ii) Parameters approximated by minimizing the total $\chi^2$ of all three data sets together, where $\chi^2_{Total}$ can be expressed as:
\begin{equation}\label{24}
\chi^{2}_{Total}= \chi^{2}_{OHD} + \chi^{2}_{Pantheon} + \chi^{2}_{BAO}
\end{equation} 
In figure 1, we have plotted the confidence contour for the derived model utilizing joint data set of OHD, Pantheon and BAO. The best estimated values of the model's free parameters by employing OHD, Pantheon, BAO and OHD + Pantheon + BAO data sets are tabulated in Table 1. It is worthwhile to note that the current model depicts small amount of anisotropy at $z = 0$ corresponds to different observation bounds. In Ref. \cite{Nojiri/2022NPB}, the authors have given a clue that the present-day expansion anisotropy provides a very informative method and could successfully be used for investigating various types of anisotropic cosmological models and their underlying theories.

%%%%%%%%%%%%%%%%%%%%%%%%%%%%%%%% Fig 1 %%%%%%%%%%%%%%%%%%%%%%%%%%%%%%%%%%%%%%%%
\begin{figure}
	\centering
	\includegraphics[scale=0.45]{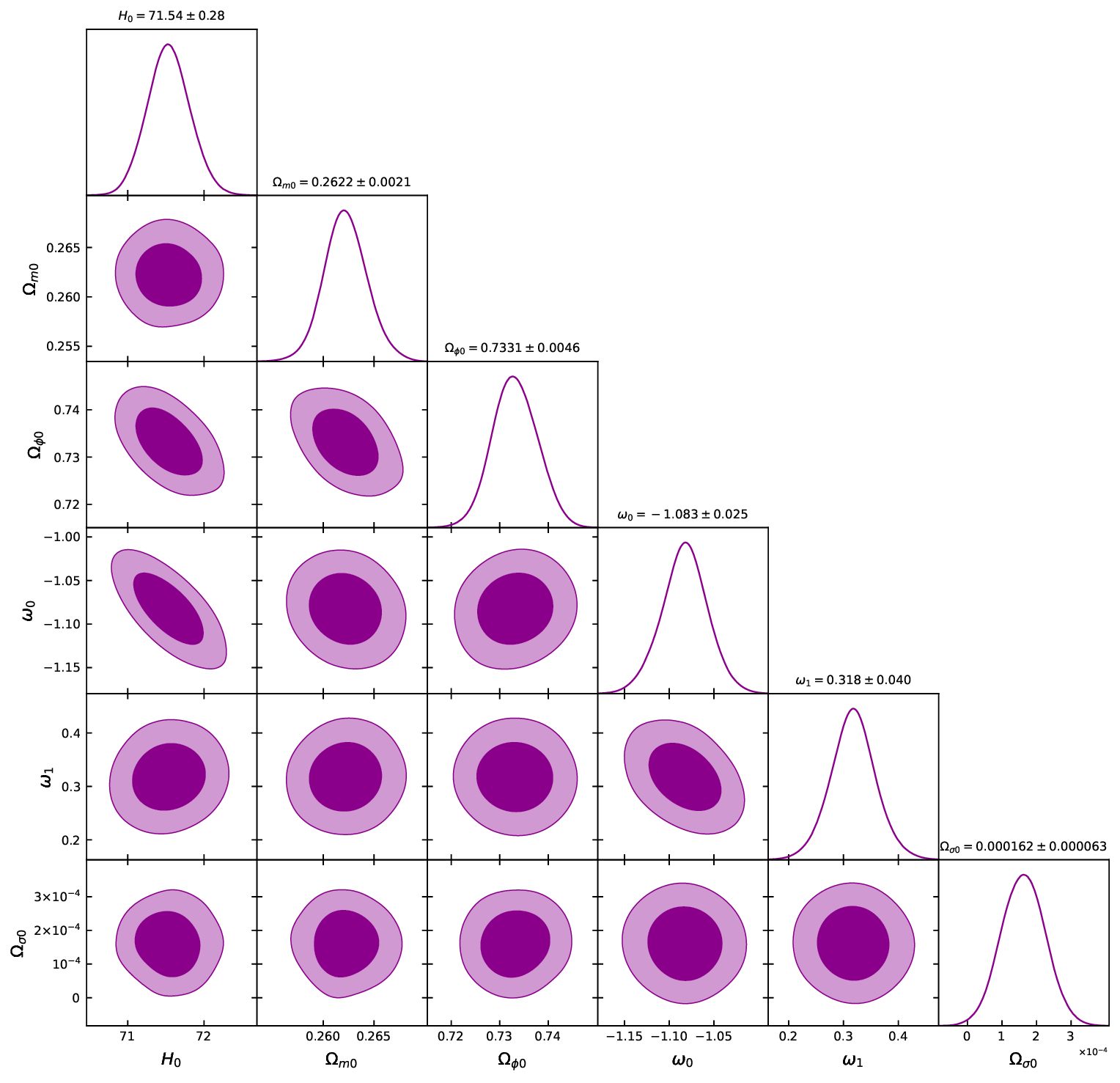}
	\caption{Plot of Confidence contour for combined set of OHD, Pantheon and BAO}. 
\end{figure}
%%%%%%%%%%%%%%%%%%%%%%%%%%%%%%%%%%%%%%%%%%%%%%%%%%%%%%%%%%%%%%%%%%%%%%%%

%%%%%%%%%%%%%%%%%%%%%%%%%%%%%%%%% Fig 2 %%%%%%%%%%%%%%%%%%%%%%%%%%%%%%%%%%%%%%%
\begin{figure}
	\centering
	(a)\includegraphics[width=5.5cm,height=5.0cm,angle=0]{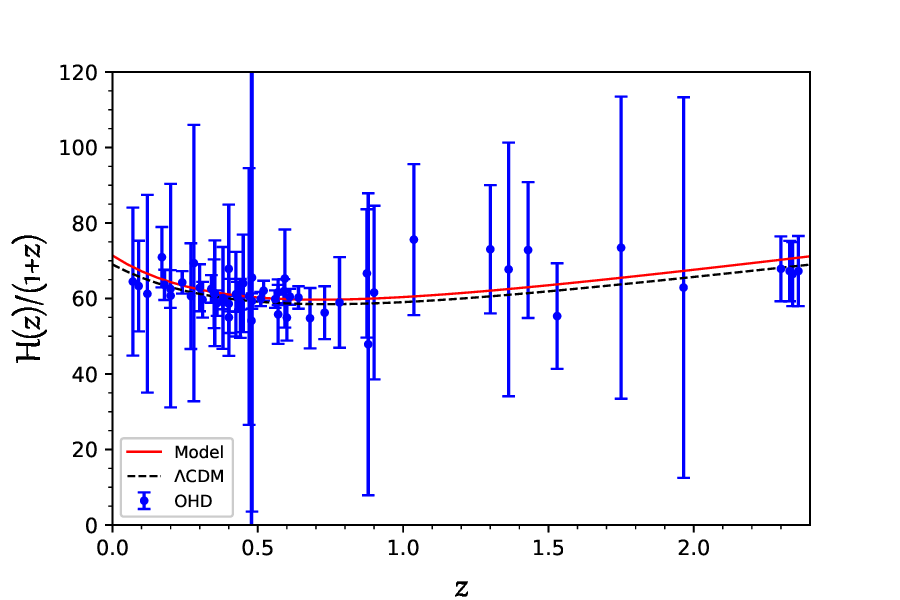}
	(b) \includegraphics[width=5.5cm,height=5.0cm,angle=0]{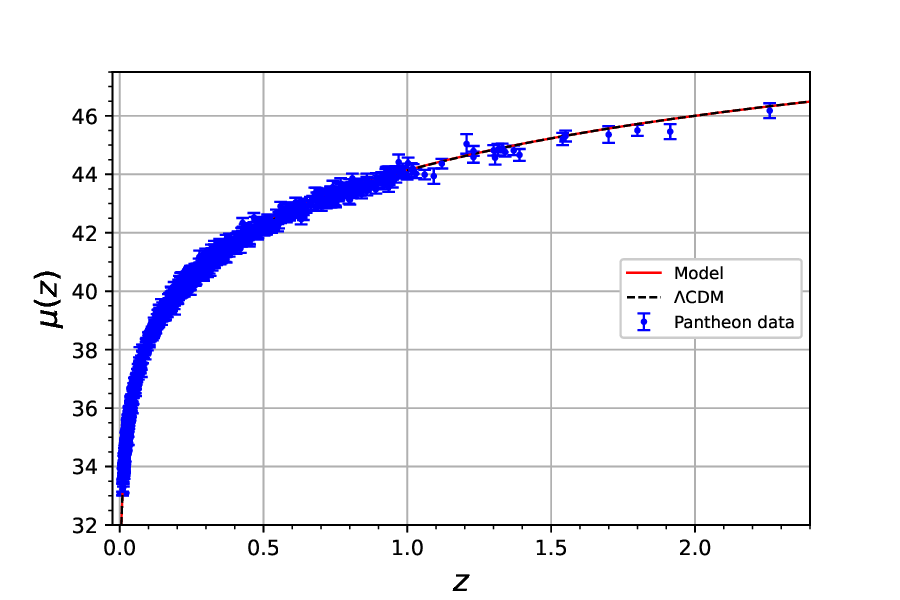}
	\caption{(a) Error-bar plot of OHD , (b) Plot of Distance modulus $\mu$ versus $z$}. 
\end{figure}
%%%%%%%%%%%%%%%%%%%%%%%%%%%%%%%%%%%%%%%%%%%%%%%%%%%%%%%%%%%%%%%%%%%%%%%%%%%%%%%%
%%%%%%%%%%%%%%%%%%%%%%%%%%%%%%%%%%%%%% Table 1 %%%%%%%%%%%%%%%%%%%%%%%%%%%%%%%%%
\begin{table}
	\caption{ The estimated values of the model's parameters for various observational datasets.}
	\begin{center}
		\begin{tabular}{|c|c|c|c|c|c|c|}
			\hline 
			\tiny Parameters & \tiny $H_{0}$ & \tiny $\Omega_{m0}$	& \tiny $\Omega_{\phi0}$ & \tiny $\omega_{0}$ & \tiny $\omega_{1}$ & \tiny $\Omega_{\sigma 0}$  \\
			\hline
			\tiny OHD & \tiny $68.7\pm1.3$ & \tiny $0.2668 \pm 0.0060$  & \tiny $0.718\pm 0.010$ & \tiny $-0.996\pm0.079$ & \tiny $0.43\pm 0.13$ & \tiny $0.00014\pm 0.00013$   \\ 
			\hline
			\tiny BAO & \tiny $67.3\pm 0.31$ & \tiny $0.262\pm 0.0020$  & \tiny $0.7299\pm0.0073$ & \tiny $-1.54\pm0.12$ & \tiny $0.38\pm0.12$ & \tiny $0.00012\pm0.00013$   \\ 
			\hline
			\tiny OHD+Pantheon & \tiny $71.82\pm0.42$ & \tiny $0.2646\pm0.0042$  & \tiny $0.7225\pm0.0095$ & \tiny $-1.094\pm0.035$ & \tiny $0.39\pm0.10$ & \tiny $0.00019\pm0.000089$   \\
			\hline
			\tiny OHD+BAO & \tiny $68.25\pm0.90$ & \tiny $0.2634\pm0.0033$  & \tiny $0.731\pm0.0061$ & \tiny $-1.026\pm0.054$ & \tiny $0.321\pm0.051$ & \tiny $0.00022\pm0.00011$   \\
			\hline
			\tiny OHD+Pantheon+BAO & \tiny $71.54\pm 0.28$ & \tiny $0.2622\pm0.0021$  & \tiny $0.7331\pm0.0046$ & \tiny $-1.083\pm 0.025$ & \tiny $0.318\pm0.040$ & \tiny $0.000162\pm0.000063$    \\
			\hline
		\end{tabular}
	\end{center}
\end{table}
%%%%%%%%%%%%%%%%%%%%%%%%%%%%%%%%%%%%%%%%%%%%%%%%%%%%%%%%%%%%%%%%%%%%%%%%%%%%%%%%%%%%%%%%%%%%
The behavior of derived model is analyzed and presented in Figure 2, with existing standard model and observational data. The comparative behavior of Hubble rate $H(z)/(1+z)$ and standard $\Lambda$CDM is plotted in Figure 2(a). In Figure 2(a), dashed black line denotes results of the standard $\Lambda$CDM model, the blue error bars represent 57 data points of OHD and solid red line shows nature of our proposed model for best fitted values of model parameters taken from combined data set of OHD, Pantheon, and BAO. For similar purpose, distance modulus $\mu(z)$ against $z$ is plotted in Figure 2(b). The distance modulus $\mu (z)$ is numerically equal to $25 + 5log10(dL/Mpc)$, where $dL = a_{0} (1+z) r = (1+z) \int_{0}^{z}\frac{dz}{H(z)}$, is the luminosity distance, that describes difference between the "apparent and absolute magnitude" of the observed supernova. Here in Figure 2(b), blue error bar represents points of considered Pantheon data as discussed earlier, dashed black line denotes results generated from the standard $\Lambda$CDM model where as solid red line shows nature of our proposed model for best fitted values of model parameters taken from combined data set of OHD, Pantheon, and BAO. From both the Figures 2(a) and 2(b) it is clearly observed that the derived model is in nice agreement with standard $\Lambda$CDM model and follows  OHD and Pantheon observations with great understanding \cite{ref47,ref54,ref55,ref56,ref56a,ref56b,ref56c}.    
%%%%%%%%%%%%%%%%%%%%%%%%%%%%%%%%%%%%%%%%%%%%
\section{Features of the model}
In this section, we depicted some evolutionary features of the scalar field model in a locally rotationally symmetric universe.
%In this section, we will discuss the features of proposed  From this model scalar field and other parameters regarding evolution of universe.

%%%%%%%%%%%%%%%%%%%%%%%%%%%%%%%%% Fig 3 %%%%%%%%%%%%%%%%%%%%%%%%%%%%%%%%%%%%%%%
\begin{figure}
	\centering
	(a)\includegraphics[width=5.5cm,height=5cm,angle=0]{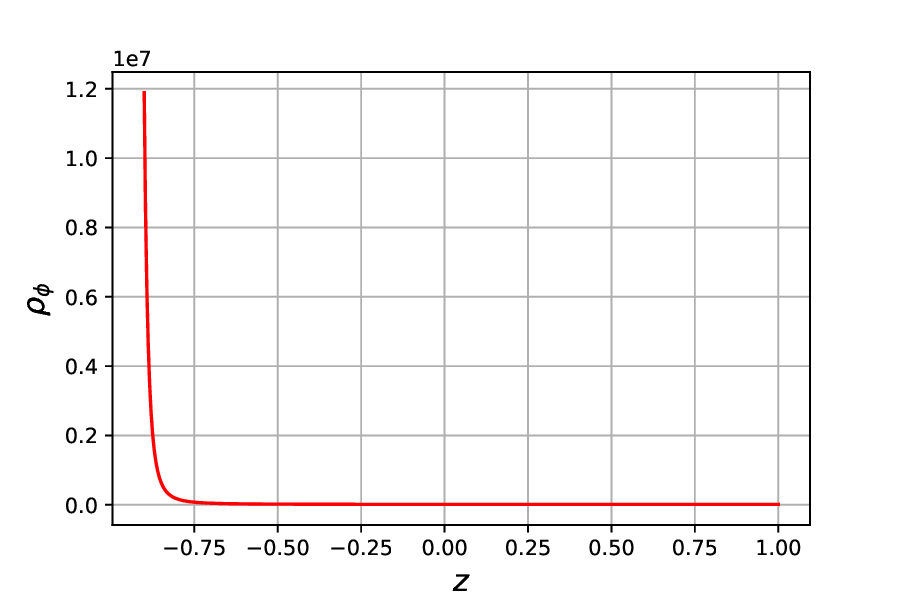}
	(b) \includegraphics[width=5.5cm,height=5cm,angle=0]{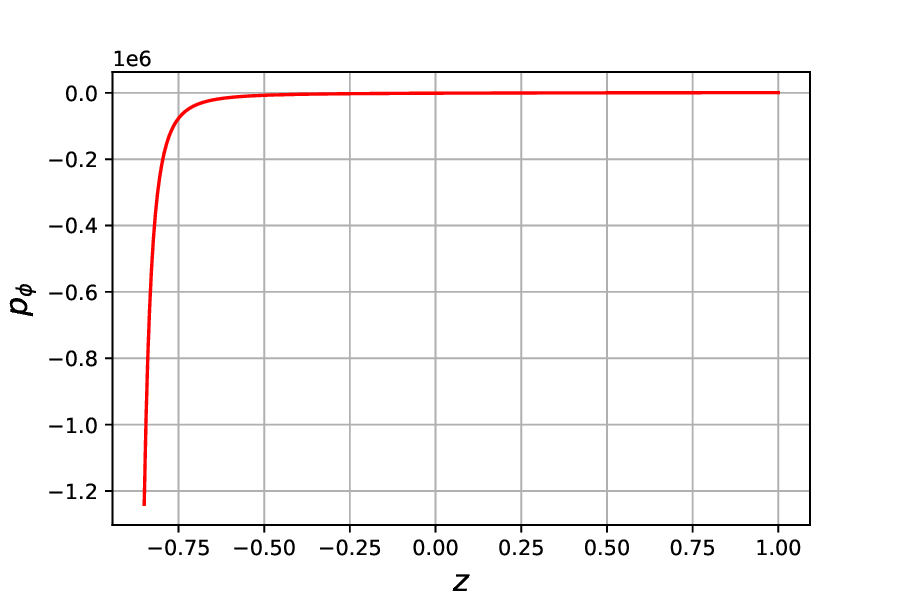}
	\caption{(a) plot of density , (b) Plot of pressure}. 
\end{figure}
%%%%%%%%%%%%%%%%%%%%%%%%%%%%%%%%%%%%%%%%%%%%%%%%%%%%%%%%%%%%%%%%%%%%%%%%%%%%%%%%
%The energy density of the proposed model is a positive increasing function of $z$ while pressure for the scalar field model remains negative during entire evolution of the universe. The negative behavior of cosmic pressure for the derived model expresses the present era of expanding universe which was dominated by dark matter in the past. It explains the fact that the space volume increase with the decrease of energy density, and provides an empty space for the current expansion. The trajectories of cosmic pressure and energy for the proposed model taking a combined data set of OHD, Pantheon, and BAO are plotted in Figure 3. 
The energy density of the proposed model is a positive increasing function of $z$ while pressure for the scalar field model remains negative during evolution of universe. The nature of pressure for scalar field confirms the universe's current expansion. The trajectories of pressure and energy density for the derived model taking a combined data set of OHD, Pantheon, and BAO are plotted in Figure 3.
%%%%%%%%%%%%%%%%%%%%%%%%%%%%%%%%%% Fig 4 %%%%%%%%%%%%%%%%%%%%%%%%%%%%%%%%%%%%%%
\begin{figure}
	\centering
	(a)\includegraphics[width=5.5cm,height=5cm,angle=0]{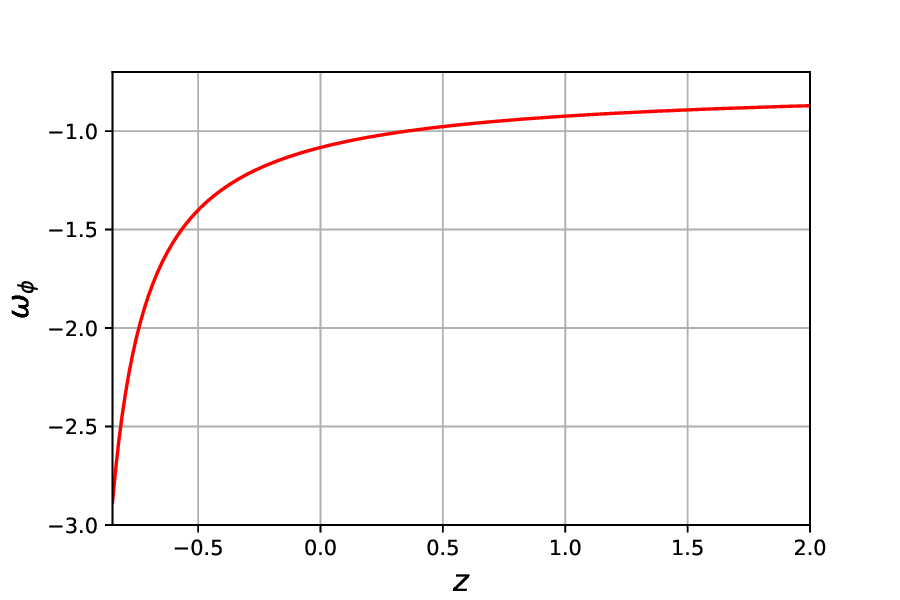}
	(b) \includegraphics[width=5.5cm,height=5cm,angle=0]{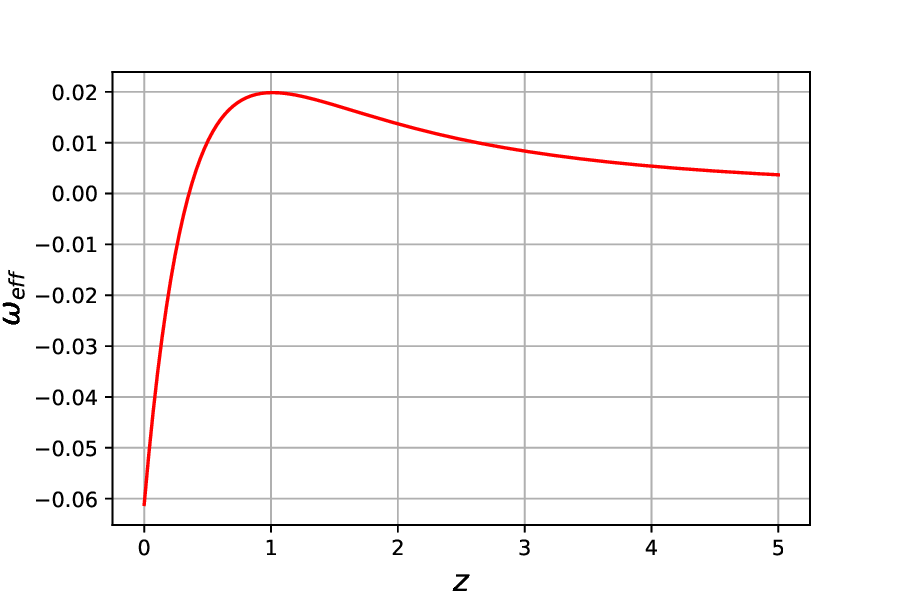}
	\caption{(a) Plot of EoS parameter for scalar field , (b) Plot of effective EoS parameter}. 
\end{figure}
%%%%%%%%%%%%%%%%%%%%%%%%%%%%%%%%%%%%%%%%%%%%%%%%%%%%%%%%%%%%%%%%%%%%%%%%%%%%%%%%
With the help of behavioural study of EoS parameter, one can analysed the various eras of the universe's cosmic evolution. For the acceleration universe, the value of $\omega$ should be $\omega < -\frac{1}{3}$. In case of $\omega = -1$, EoS parameter approaches to the $\Lambda$CDM model. The model lies the quintessence era for $-1< \omega \leq 0$ and a phantom scenario is depicted for $\omega < -1$. $\omega  = 0$ signifies matter influenced scenario and  $\omega = \frac{1}{3}$ express radiation dominated universe. It has been observed that the EoS parameter for the derived scalar field model lies in the phantom era for combined data set. The effective EoS parameter of the proposed model lies in quintessence era that refers to a period in the history of the universe where dominant form of DE is believed to be a scalar field, which is a function of time. The nature of EoS for the present model is plotted in Figure 4.
%%%%%%%%%%%%%%%%%%%%%%%%%%%%%%%%%% Fig 5 %%%%%%%%%%%%%%%%%%%%%%%%%%%%%%%%%%%%%%
\begin{figure}
	\centering
	(a)\includegraphics[width=5.5cm,height=5cm,angle=0]{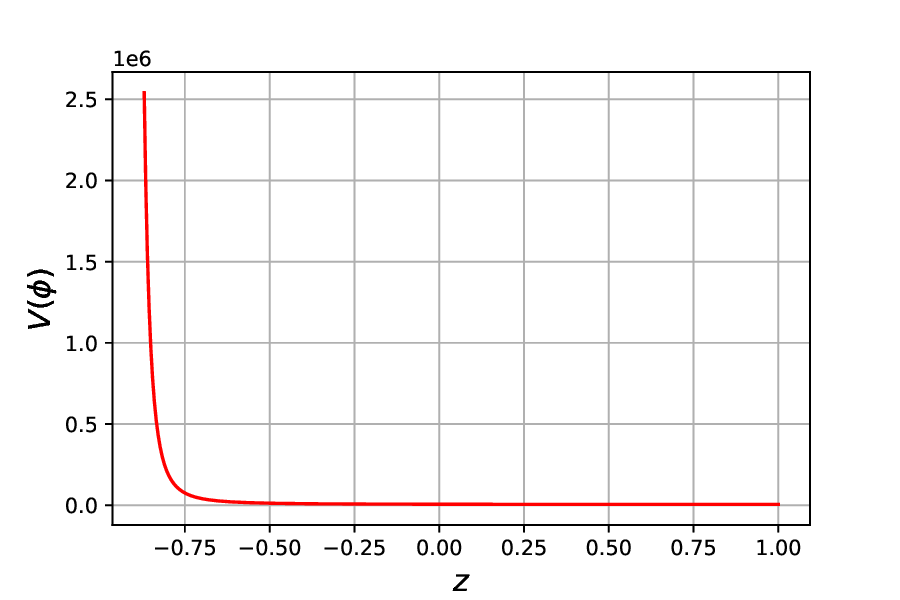}
	(b) \includegraphics[width=5.5cm,height=5cm,angle=0]{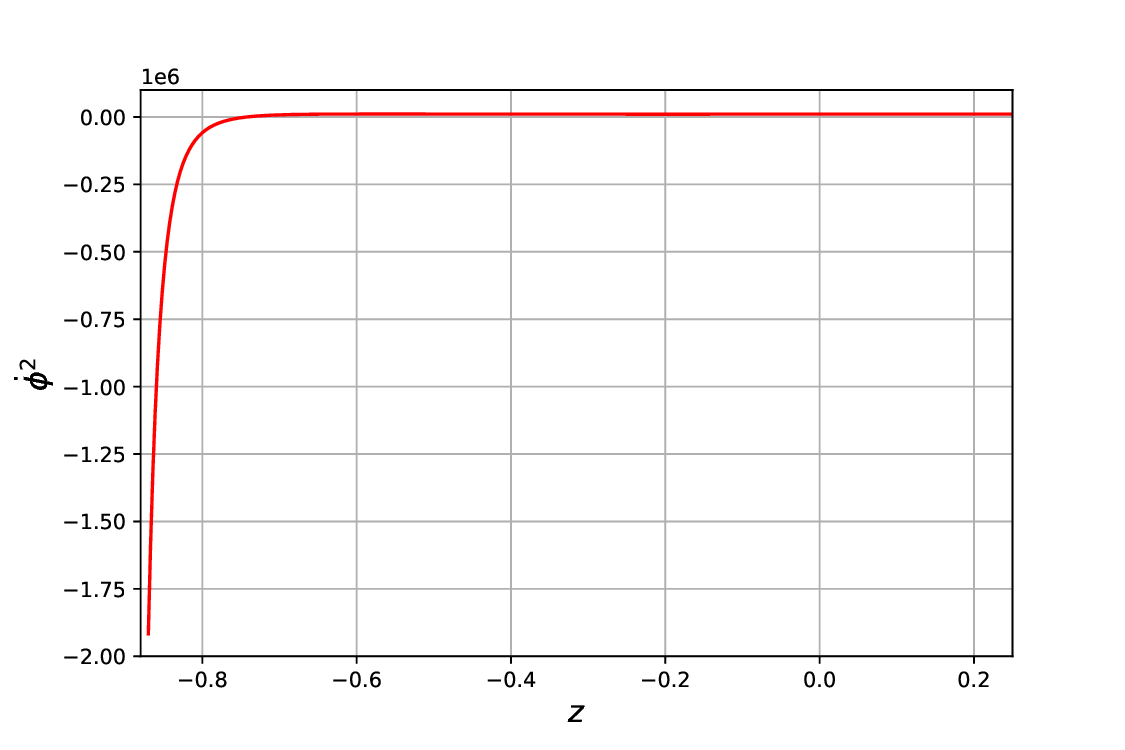}
	\caption{(a) plot of potential , (b) Plot of scalar field}. 
\end{figure}
%%%%%%%%%%%%%%%%%%%%%%%%%%%%%%%%%%%%%%%%%%%%%%%%%%%%%%%%%%%%%%%%%%%%%%%%%%%%%%%%
The behavior of scalar field and its potential are determined by the physical system taken into consideration. In physics, negative energy can possess non-physical or unstable solutions while a positive scalar potential is normally linked with stable formations. However, a negative scalar potential might be physically applicable in some models of dark energy or inflation in cosmology. Figure 5 (b) shows that the negative kinetic energy $\frac{1}{2}\dot{\phi}^2$  has a potential $V(\phi)$  that causes a late time expansion of universe. The kinetic term $\dot{\phi}^2$ as the contributor of dark energy depict the current expansion of cosmos as shown by negative trajectory of  $\dot{\phi}^2$ in figure 5(b)\cite{ref57,ref58,ref59}.
%%%%%%%%%%%%%%%%%%%%%%%%%%%%%%%%%%%%%%%%
\subsection{Age of Universe}
In previous section we have discussed the dependency of scale factor ($a$) on redshift ($z$). These relations can be used to calculate age of the cosmos as
\begin{equation}\label{25}
dt=-\frac{dz}{(1+z) H(z)}\implies \int_{t}^{t_{0}} dt=-\int_{z}^{0} \frac{1}{(1+z) H(z)} dz 
\end{equation}
Using Eq.($16$) and value of $H(z)$ of proposed model, the present age of universe $t_0$ can be calculated as
\begin{equation}\label{26}
t_{0}=\lim_{x \to \infty}\int_0^x \frac{1}{H_{0} (z+1) \sqrt{\Omega_{m0} (1+z)^{3}+\Omega_{de0} (1+z)^{3(1+\omega_{de})}+\Omega_{\phi0} (1+z)^{6}+\Omega_{\sigma0} (1+z)^{6}}} \, dz
\end{equation}
The age of universe can be extracted from above equation whose solution can be obtained by plotting $ H_{0} (t_{0}-t) $ as a redshift function. 
The $ H_{0} (t_{0}-t) $ is plotted for OHD data and presented in Figure 6(a). Utilizing Figure 6(a), we obtained $H_{0} \ t_{0} = 0.97786$, that leads to present age $ t_{0}=13.79 $ $ Gyrs $. The calculated age of the universe through proposed model for OHD, Pantheon, BAO and combination of these three data is tabulated in Table 2. The proposed age of universe is in confirmation with results obtained in recent observations as suggested by \cite{ref56,ref61,ref62,ref63}.
%%%%%%%%%%%%%%%%%%%%%%%%%%%%%%%%%% Fig 6 %%%%%%%%%%%%%%%%%%%%%%%%%%%%%%%%%%%%%%
\begin{figure}
	\centering
	(a)\includegraphics[width=5.5cm,height=5cm,angle=0]{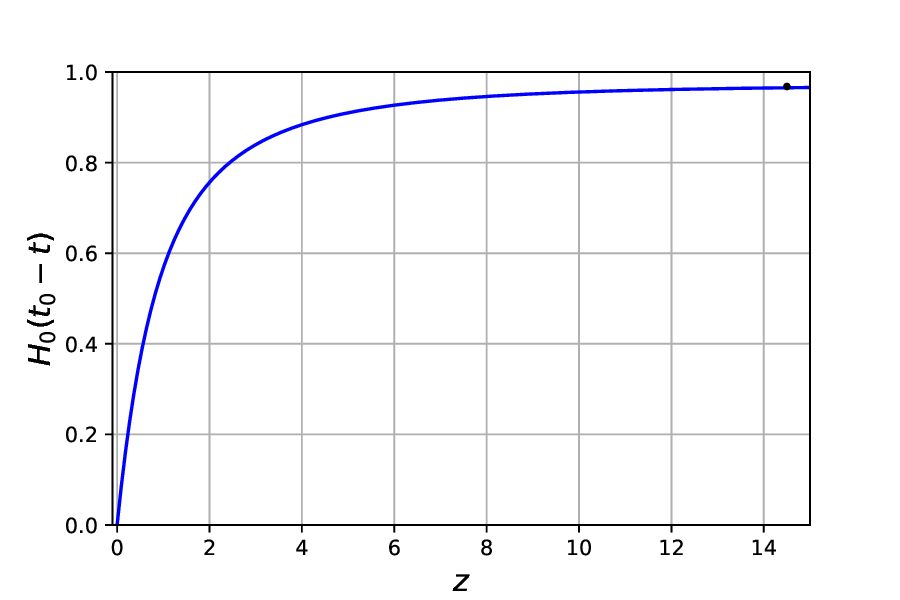}
	(b) \includegraphics[width=5.5cm,height=5cm,angle=0]{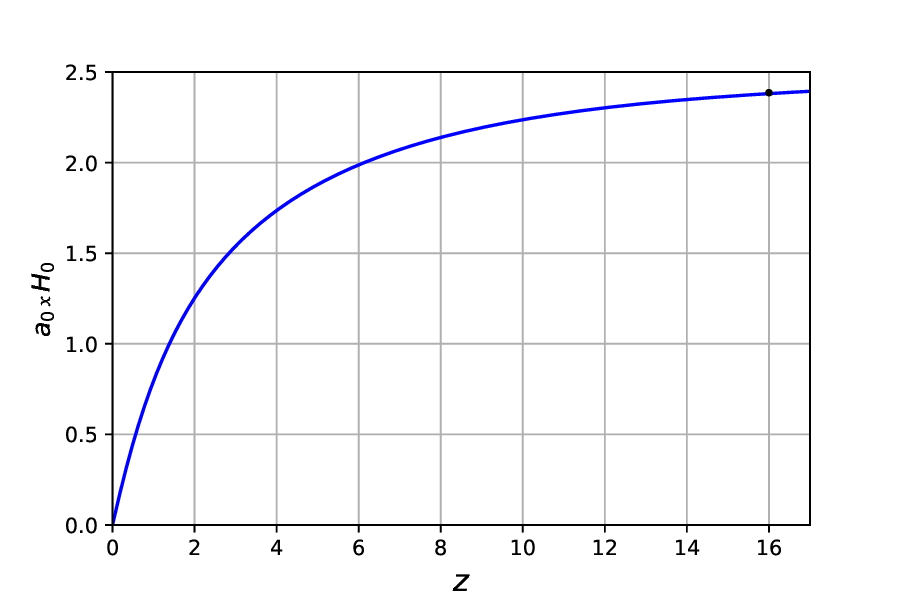}
	\caption{(a) Plot of $ H_{0} (t_{0}-t) $ versus $ z $, (b) Plot of particle horizon $q$ vs $z$}. 
\end{figure}
%%%%%%%%%%%%%%%%%%%%%%%%%%%%%%%%%%%%%%%%%%%%%%%%%%%%%%%%%%%%%%%%%%%%%%%%%%%%%%%%
\subsection{Particle horizon}
In cosmology, particle horizon is an important parameter that relates universe expansion and it's observable boundaries. A light beam emitted from a source situated at particle horizon have travelled from the beginning of the expansion to the observer through the age of universe \cite{ref64}. That means it is the maximum distance from where information of universe beginning can be retrieved. Since universe has gone through expansions of different rate, it is appropriate to express it in terms of proper length. Suppose a light beam from a source is emitted at time $t_p$ travelling along $x$ direction in expanding universe whose scale factor is $a_{0}$, reaches the observer at $t_{0}$ (present time) then the proper distance of the source from observer can be expressed as 
$r = a_{0} \int_{t_p}^{t_0} {\frac{dt}{a(t)}}$.\\
The particle horizon $R_p$ will be equals to $r$ in the limit ${t_{p} \to 0}$.  Thus $R_p$ can be defined as $\lim_{t_{p} \to 0} a_{0} \int_{t_p}^{t_0} {\frac{dt}{a(t)}} = \lim_{z \to \infty} \int_{0}^{z}{\frac{dz}{H(z)}}$. Using Eq. (12), the expression of particle horizon as a function of $z$ can be expanded as:
\begin{equation}\label{27}
R_{p}=\lim_{x \to \infty}\int_0^x \frac{1}{H_{0} \sqrt{\Omega_{m0} (1+z)^{3}+\Omega_{de0} (1+z)^{3(1+\omega_{de})}+\Omega_{\phi0} (1+z)^{6}+\Omega_{\sigma0} (1+z)^{6}}} \, dz
\end{equation}
The behaviour of particle horizon parameter as function of $z$ for combined data (OHD + Pantheon + BAO) is plotted in Figure 6(b). The analysis of proposed model reveals $R_p \to 2.386 H_{0}^{-1}$ at red shift $10^4$ as shown in Figure 6(b). Here we have used $H_0=71.25$ $km/s/Mpc$ $\sim 0.0727$ $Gyrs^{-1}$. The estimation of $R_p$ for different set of observational data is tabulated in Table 2. The observed theoretical values of $Rp$ as revealed by proposed model are in great agreement with the result obtained in \cite{ref37}. The proper distance $x$ tends to infinite at $z = 0$ as clearly noticed from the Figure 6(b). As a result, the event that occurred at the origin of universe is very far away from us (or at infinite distance). 
%%%%%%%%%%%%%%%%%%%%%%%%%%%%%%%%%%%%%%%%%%%%%%%%%%%%%%%%%%%%%%%%%%%%%%%%%%%%%%%%

\subsection{Deceleration parameter}
Deceleration parameter (DP) is one of the crucial parameters that depicts the phase transition of the cosmos among the other parameters that disclose the dynamics of the universe's evolution. The DP can be defined as as $q= -\frac{\ddot a}{a H^2} = -1 + \frac{(1+z)}{H(z)} \frac{d H(z)}{dz} $. For the derived model, the expression for DP is developed as given below. 
\begin{equation}\label{28}
q = \frac{(3 \omega_{de} +1) \Omega_{de0} (z+1)^{3 \omega_{de} }+\Omega_{m0}+4 (z+1)^3 \Omega_{\sigma 0}+4 (z+1)^3 \Omega_{\phi 0}}{2 \left(\Omega_{de0} (z+1)^{3 \omega_{de}}+\Omega_{m0}+(z+1)^3 \Omega_{\sigma 0}+(z+1)^3 \Omega_{\phi 0}\right)}
\end{equation}
Main features of expanding cosmos can be easily explained by the analysis of DP and Hubble parameters. The universe's phase transition and Age can be described accurately from DP and Hubble parameter.\\

The DP for the derived model is computed for the best fitted model parameters estimated taking OHD, Pantheon, BAO, and their combination into account. The calculated $q_{0}$ values for distinct data sets are tabulated in Table 2. We have shown the signature flipping nature of the cosmos through graphical presentation of DP in Figure 7. The transitioning characteristic of the cosmos indicates the presence of dark energy. The transitioning of cosmos from deceleration phase to present accelerated expansion era of universe with transition redshift at $z_{t}=0.695$ is presented in Figure 7. The results derived for the suggested model agree well with recent experimental findings \cite{ref7,ref37,ref53,ref67,ref68,ref69,ref70,ref71,ref72,ref73,ref74}.

%%%%%%%%%%%%%%%%%%%%%%%%%%%%%%%%%%% Figure 7 %%%%%%%%%%%%%%%%%%%%%%%%%%%%%%%%%
\begin{figure}
	\centering
	\includegraphics[scale=0.5]{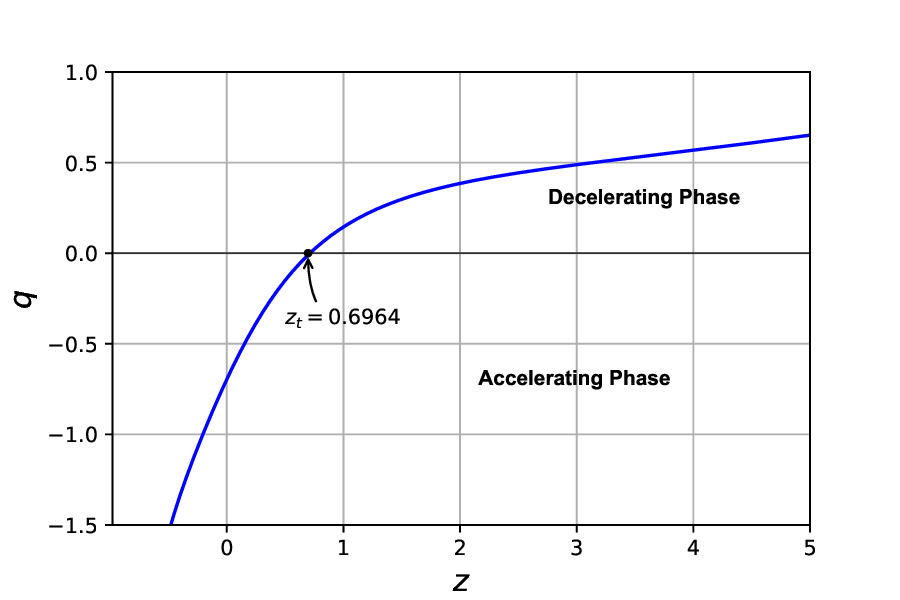}
	\caption{Plot of deceleration parameter $q$ vs $z$}\label{fig7}	
\end{figure}
%%%%%%%%%%%%%%%%%%%%%%%%%%%%%%%%%%%%%%%%%%%%%%%%%%%%%%%%%%%%%%%%%%%%%%%%%%%%%%%%%%%%%%%%%%
\subsection{Jerk Parameter}
Another diagnostic tool that is widely used in cosmology is the jerk parameter ($j$). The thought that there should be a shock that causes the universe to shift from a decelerating phase to accelerating scenario gave rise to the idea. In cosmological studies, jerk parameter is the third order term in Taylor's expansion of $a$ about $a_0$. It gives us an extra edge in identifying kinematically degenerated cosmological models \cite{ref75}. Due to involvement of scale factor's third derivative, it is more accurate than the Hubble parameter. For the proposed model, the cosmic jerk parameter is defined as \cite{ref75,ref76}
\begin{equation}\label{29}
j=1-(1+z) \frac{H'(z)}{H(z)}+\frac{1}{2}(1+z)^2 \left[\frac{H''(z)}{H(z)}\right]^2
\end{equation}
Here, $H'(z)$ and $H''(z)$ are the derivatives of $H(z)$ with $z$. Behaviour of jerk parameter $j$ for the best fit parameters of proposed model taking OHD + BAO + Pantheon combinedly into account is presented in Figure 8(a).

Figure 8(a) shows that the developed models evolve with positive jerk parameter values other than one. In the models under discussion, $q$ is negative whereas $j\neq1$. As a result, the suggested solution defined a universe model other than the $\Lambda$CDM. The positive behavior of jerk measure verify the faster expansion of cosmos in late-time \cite{ref77,ref78,ref79}.
%The late time acceleration is confirmed by the positive value of the jerk measure \cite{ref77,ref78,ref79}.

%%%%%%%%%%%%%%%%%%%%%%%%%%%%%%%%% Fig 8 %%%%%%%%%%%%%%%%%%%%%%%%%%%%%%%%%%%%%%
\begin{figure}
	\centering
	(a)\includegraphics[width=5.5cm,height=5cm,angle=0]{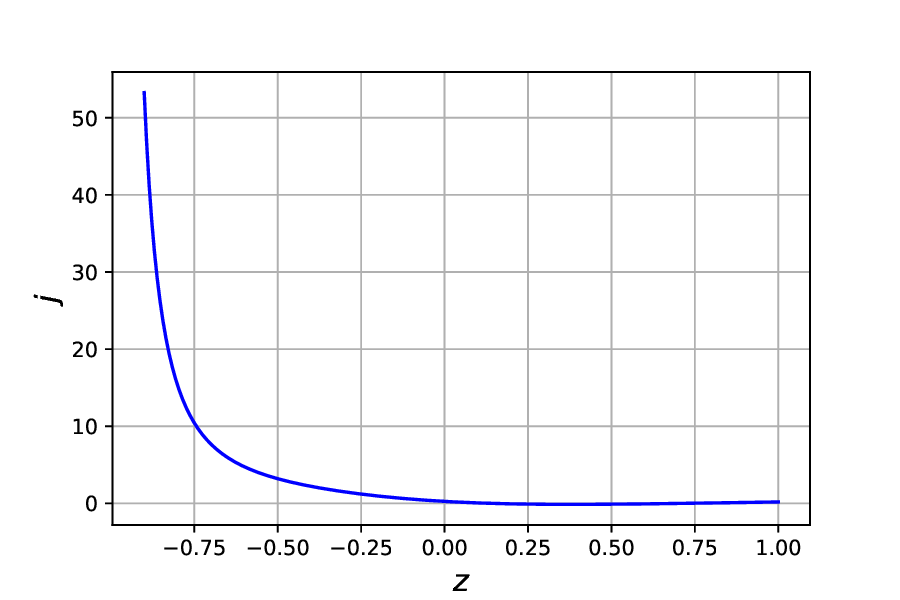}
	(b) \includegraphics[width=5.5cm,height=5cm,angle=0]{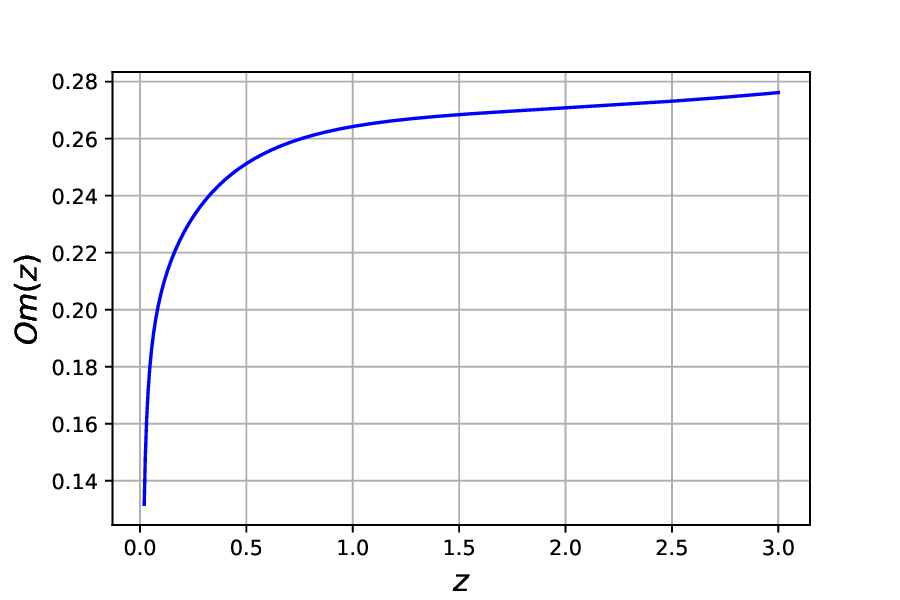}
	\caption{(a) Plot of jerk parameter $ j $ versus $ z $, (b) Evaluation of $ Om(z) $ versus $ z $}. 
\end{figure}
%%%%%%%%%%%%%%%%%%%%%%%%%%%%%%%%%%%%%%%%%%%%%%%%%%%%%%%%%%%%%%%%%%%%%%%%%%%%%%%%

\subsection{Om diagnostic}
The $O_m$ diagnostic analysis is a useful geometrical diagnostic proposed by Sahni et al. \cite{ref80} as a complement of the statefinder diagnostics. The $O_m$ diagnostic is a function of Hubble and assumed to be the simplest diagnostic tool \cite{ref81}. In literature, many cosmological models have been proposed to resolve the mystery of DE. $O_m$ parameter is an efficient tool to differentiate between these DE model \cite{ref78}. For the negative trajectory of the $O_m$ tool depicts the quintessence behaviour $(\omega > -1)$ of the model and for its positive trajectory model behaves like phantom $(\omega < -1)$. $O_m$ parameter is defined as $O_{m}(z) = \frac{E^2 (z) - 1}{(1+z)^3 -1}$, where $E(z) = \frac{H(z)}{H_0}$ with $H_0$  present value of Hubble constant.\\
For the proposed model the $O_m$ diagnostics is found as

\begin{equation}\label{32}
Om(z)=\frac{\Omega_{m 0} (z+1)^3 +\Omega_{\phi 0} (z+1)^{3 \left(\omega_{0}+\omega_{1}+1\right)} e^{-\frac{3 \left(\omega _1 z\right)}{z+1}}+\Omega_{\sigma 0} (z+1)^6-1}{z \left(z^2+3 z+3\right)}
\end{equation}
The variation of $O_{m}(z)$ parameter versus $z$ is depicted in Figure 8 (b) for best estimated values of parameter. The positive curvature discriminates the proposed cosmic model  with standard  $\Lambda$CDM model. Thus, the proposed model shows a behavior of phantom like universe.
%%%%%%%%%%%%%%%%%%%%%%%%%%%%%%%%%%%%%%%% Table 2 %%%%%%%%%%%%%%%%%%%%%%%%%%%%%%%%%%%%%%%%%
\begin{table}
	\caption{ Summary of derived results of the proposed model  }
	\begin{center}
		\begin{tabular}{|c|c|c|c|c|c|}
			\hline 
			\tiny Parameters & \tiny $q_{0}$ & \tiny $z_{t}$	& \tiny Age ($t_{0}$) Gyrs & \tiny $j_{0}$ & \tiny $R_p $   \\
			\hline
			\tiny OHD & \tiny $-0.5888\pm0.089$ & \tiny $0.65^{+0.084}_{-0.083}$  & \tiny $13.485^{+0.907}_{-0.822}$ & 
			\tiny $0.0848^{+0.104}_{-0.111}$ & \tiny $2.345^{+0.194}_{-0.147}$.$H^{-1}_{0}$    \\ 
			\hline
			\tiny BAO & \tiny $-1.1995\pm0.132$ & \tiny $0.635^{+0.017}_{-0.020}$  & \tiny $15.04^{+0.538}_{-0.511}$ & 
			\tiny $1.6407^{+0.531}_{-0.463}$ & \tiny $2.532^{+0.183}_{-0.127}$.$H^{-1}_{0}$ \\ 
			\hline
			\tiny OHD+Pantheon & \tiny $-0.70^{+0.039}_{-0.041}$ & \tiny $0.680^{+0.018}_{-0.028}$  & \tiny $13.17^{+0.427}_{-0.407}$ & \tiny $0.2602^{+0.076}_{-0.069}$ & \tiny $2.360^{+0.105}_{-0.080}$.$H^{-1}_{0}$ \\
			\hline
			\tiny OHD+BAO & \tiny $-0.6307^{+0.060}_{-0.061}$ & \tiny $0.691^{+0.025}_{-0.026}$  & \tiny $13.69^{+0.532}_{-0.546}$ & \tiny $0.1462^{+0.099}_{-0.085}$ & \tiny $2.325^{+0.111}_{-0.088}$.$H^{-1}_{0}$ \\
			\hline
			\tiny OHD+Pantheon+BAO & \tiny $-0.6964\pm0.028$ & \tiny $0.6964^{+0.0136}_{-0.0006}$  & \tiny $13.39^{+0.116}_{-0.398}$ & \tiny $0.2537^{+0.052}_{-0.049}$ & \tiny $2.386^{+0.066}_{-0.058}$.$H^{-1}_{0}$ \\
			\hline
		\end{tabular}
	\end{center}
\end{table}
%%%%%%%%%%%%%%%%%%%%%%%%%%%%%%%%%%%%%%%%%%%%%%%%%%%%%%%%%%%%%%%%%%%%%%%%%%%%%%%%%%%%%%%%%%%%

\subsection{Energy Conditions}
%The viability of models is the major issue in theoretical modelling of cosmos in framework of modified theories. There are several prevailing energy conditions (ECs) in GR that impose certain conditions to avoid a region whose energy density being negative. These conditions proposed a generalized viability to the fact that energy density of cosmos always remains non-negative to the entire EMT \cite{ref82,ref83}. Many significant space-time, black hole, and wormhole singularity problems can be examined using ECs in a variety of contexts. The well-known Raychaudhari equation \cite{ref84} can be used to analyze the viability of the numerous ECs that are frequently utilized in GR. The most popular energy condition of GR are termed as Weak energy condition (WEC), Null energy condition (NEC), Dominant energy condition (DEC, and Strong energy condition (SEC). These ECs mathematically can be expressed as: (i)   $\rho +  p \geq 0$ , $\rho \geq 0$ (WEC), (ii) $\rho + p \geq 0$ (NEC), $\rho - p \geq 0$ (DEC), and $\rho + 3 p \geq 0$ (SEC) respectively.\\
There are several prevailing energy conditions (ECs) in GR that apply restrictions to prevent an area with a negative energy density. These circumstances suggested a generalised validity to the idea that the universe's energy density always maintains a positive relationship with the entire EMT \cite{ref82,ref83}. The viability of the several ECs that are regularly employed in GR can be examined using the well-known Raychaudhari equation \cite{ref84}. The most common energy conditions in GR can be represented mathematically as follows:
%%%%%%%%%%%%%%%%%%%%%%%%%%%%%
\begin{itemize}
	\item  WEC $ \Leftrightarrow \rho +  p \geq 0$ , $\rho \geq 0$ 
	\item  NEC $\Leftrightarrow \rho + p \geq 0$
	\item  DEC $\Leftrightarrow \rho - p \geq 0$
	\item  SEC $\Leftrightarrow \rho + 3 p \geq 0$
\end{itemize}
%%%%%%%%%%%%%%%%%%%%%%%%%%%%%%%%%%%%%%%%%%%%%%%%%%%%%% 
Given that $V(\phi)$ and $\frac{\dot{\phi}^2}{2}$ can be used to develop the pressure and energy density of the suggested model, respectively. So, in form of scalar field, the ECs of can be established as: (i) NEC: $\forall \, V(\phi)\geq0$, (ii) WEC $\Leftrightarrow V(\phi)\geq \frac{\dot{\phi}^2}{2}$, (iii) SEC $\Leftrightarrow V(\phi)\geq \dot{\phi}^2$, (iv) DEC $\Leftrightarrow V(\phi)\geq0$. According to the WEC, the universe's matter energy density is always positive. The DEC explains why the measured energy flux, which can never be greater than the speed of light, equals the experimental energy density, which remains positive.  The WEC and DEC are always satisfied by all recognised energy and matters. The unusual energy (DE) that creates a strong negative pressure is responsible for the universe's fast expansion and the violet SEC. The NEC violation is more exotic and implies the presence of ``exotic matter," which is sometimes referred to as matter with a negative energy density \cite{ref85,ref86}.
%%%%%%%%%%%%%%%%%%%%%%%%%%%%%%%%%%% Figure 9 %%%%%%%%%%%%%%%%%%%%%%%%%%%%%%%%%
\begin{figure}
	\centering
	\includegraphics[scale=0.6]{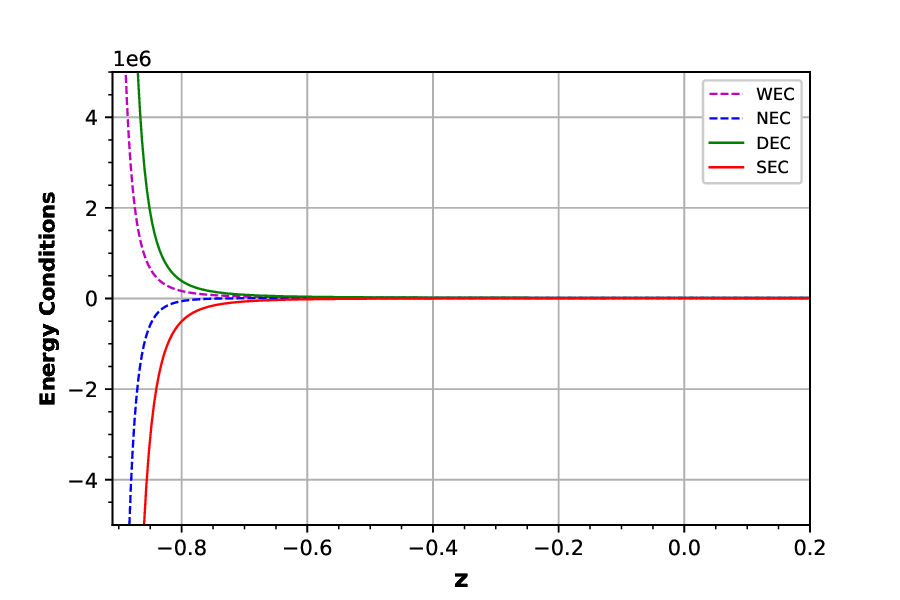}
	\caption{Plot of energy conditions }\label{fig15}	
\end{figure}
%%%%%%%%%%%%%%%%%%%%%%%%%%%%%%%%%%%%%%%%%%%%%%%%%%%%%%%%%%%%%%%%%%%%%%%%%%%%%%%%%%%%%%%%%%
%For the combined observational data sets of OHD, Pantheon, and BAO, all energy conditions for the derived model are demonstrated in Figure 9. WEC and DEC for the specified model are satisfied with the combined observational data set, as seen in Fig. 9. The violation of SEC for the derived model depicts the accelerated expansion of universe and evidence of exotic matter in the cosmos \cite{ref87,ref88}. For $\rho_{\phi}+3 p_{\phi} < 0$, the model lies in inflation era which indicates $V(\phi) > \frac{\dot{\phi}^2}{2}$ \cite{ref19,ref20}. 

For combined dataset (BAO+OHD+Pantheon), all energy conditions for the resulting model are shown in Figure 9. For the derived model, WEC and NEC are satisfied while NEC and SEC are violated. The violation of SEC depicts the accelerated expansion of universe and existence of exotic matter in the universe \cite{ref87,ref88}. The model lies in expansion era of universe for $\rho_{\phi}+3 p_{\phi} < 0$ which indicates $V(\phi) > \frac{\dot{\phi}^2}{2}$ \cite{ref19,ref20}. The violation of NEC suggested, the conditions of a bouncing scenario in the derived model. From the evolutionary behaviour of the EoS parameter in figure 9, it can be observed that the NEC and SEC are violated when the universe crosses the phantom line and enters the phantom phase, and model behave like quintom model.

\section{Conclusion}
In this study, we have explored a transitioning cosmological model in the axially symmetric Bianchi type-I universe the background of scalar field theory. The model parameters are estimated taking observational datasets of BAO, Pantheon, and OHD using MCMC analysis. Some dynamical properties like potential, jerk parameter are described. The important characteristics of the proposed model are given below  

\begin{itemize}
	\item[i)] The confidence contour plot of model parameter for joint data set of OHD, BAO, and Pantheon is presented in figure 1. The best estimated values of the model's free parameters by employing OHD, Pantheon, BAO and their combined data sets are tabulated in Table 1.
	
	\item[ii)] The comparative behavior of Hubble rate $H(z)/(1+z)$ and standard $\Lambda$CDM is plotted in Figure 2(a) and distance modulus $\mu(z)$ against $z$ in Figure 2(b). From both the Figures 2(a) and 2(b) it is clearly observed that the derived model is in nice agreement with standard $\Lambda$CDM model and follows  OHD and Pantheon observations with great understanding \cite{ref47,ref54,ref55,ref56}. 
	
	\item[iii)]The behavior of energy density for scalar field model remains positive during entire evolution of the universe while cosmic pressure of scalar field remains negative. The natures of energy density and cosmic pressure of scalar field for combined observational dataset can be seen in Fig. 3.
	
	\item[iv)] From Figure 4, it has been observed that the EoS parameter for scalar field lies in the phantom era for combined data set of OHD, BAO, and Pantheon. 
	
	\item[v)] The negative behavior of kinetic term $\dot{\phi}^2$ in figure 5(b) shows the influence of dark energy in the expansion of universe and the derived model behaves like phantom model of universe \cite{ref57,ref58,ref59}. 
	
	\item[vi)] The age for derived model is found to be 13.79 $Gyrs$ that agree with the recent findings \cite{ref56,ref61,ref62,ref63}. 
	
	\item[vii)] The observed theoretical values of particle horizon $Rp\to 2.386 H_{0}^{-1}$ as revealed by proposed model are in great agreement with the result obtained in \cite{ref37}.
	
	\item[viii)]  The shows the signature flipping nature of cosmos from deceleration phase to present accelerated expansion era of universe with transition redshift at $z_{t}=0.695$ is presented in Figure 7 which are well supported by recent experimental findings \cite{ref7,ref53,ref67,ref68,ref69,ref70,ref71,ref72,ref73,ref74}.
	
	\item[ix)] The late time acceleration is confirmed by the positive value of the jerk measure \cite{ref77,ref78,ref79}.
	
	\item[x)] The positive curvature discriminates the proposed cosmic model  with standard  $\Lambda$CDM model. The result confirms the phantom behavior of the proposed model.	
	
	\item[xi)] The violation of SEC for the derived model depicts the present expansion of universe at faster rate and indication of exotic matter in the universe \cite{ref87,ref88}.
	
\end{itemize}
For the proposed model, some derived results for distinct datasets are summarized in Table 2. The scenario of present accelerated expansion of the universe is described by contribution of scalar field. The obtained results are in excellent agreement with recent findings. 

\section*{Data availability} All data generated or analysed during this study are included in this article.

%\section*{Acknowledgment}
%\noindent The authors wish to place on record their sincere thanks to the reviewer(s) for illuminating suggestions that have significantly improved our manuscript in terms of research quality.

%%%%%%%%%%%%%%%%%%%%%%%%%%%%%%%%%%%%%%%%%%%%%%%%%	


\begin{thebibliography}{99}
	\bibitem{ref1} A. Einstein, Sitzungsber Preuss Akad Wiss Berlin (Math.Phys.) 1917 142 (1917).
	\bibitem {ref2} 
	A.G. Riess et al., Astron. J. 116, 1009 (1998)
	\bibitem {ref3}
	S. Perlmutter et al., Astrophys. J. 517, 565 (1999)
	\bibitem {ref4} 
	R.R. Caldwell et al., Phys. Rev. D {\bf 69} 103517 (2004) 
	\bibitem {ref5}
	D.N. Spergel et al., Astrophys. J. Suppl. {\bf 148}, 175 (2003)
	\bibitem {ref6}
	E. Komatsu et al., Astrophys. J. Suppl. {\bf 180} 330 (2009)
	\bibitem {ref7}
	A.G. Riess et al., { Astrophys. J.} {\bf 607} 665 (2004)
	\bibitem {ref8} 
	T. Koivisto, D.F. Mota, Phys. Rev. D {\bf 73} 083502 (2006) 
	\bibitem {ref9} 
	P.J.E. Peebles, B. Ratra, Rev. Mod. Phys. {\bf 75} 559 (2003)
	\bibitem{Bamba/2012}
	K. Bamba, S. Capozziello, S. Nojiri, S. D. Odintsov, Astrophys.Space Sci. {\bf 342} 155-228 (2012)
	
	\bibitem{Akarsu/2010} \"{O} Akarsu, C. B. Killinc, Gen. Relativ. Gravit. {\bf 42} (2010) 119.
	
	\bibitem{Kumar/2011} S. Kumar and C.P. Singh, Gen. Relativ. Gravit. {\bf 43} (2011) 1427.  
	
	\bibitem{Kumar/2011a} S. Kumar and A. K. Yadav, Mod. Phys. Lett. A {\bf 26} (2011) 647.
	
	\bibitem{Yadav/2011} A. K. Yadav, Astrophys. Space Sci. {\bf 335} (2011) 565. 
	
	\bibitem{Nojiri/2004} S. Nojiri, S. D. Odintsov, Phys. Rev. D {\bf 103522} (2004) 70.
	
	\bibitem{Nojiri/2005} S. Nojiri, S. D. Odintsov, S. Tsujikawa, Phys. Rev. D {\bf 71} (2005) 063004. 
	
	\bibitem{Nojiri/2006} S. Nojiri, S. D. Odintsov, Gen. Rel. Grav. {\bf 38} (2006) 1285. 
	
	\bibitem{Nojiri/2005a} S. Nojiri, S. D. Odintsov, Phys. Rev. D {\bf 72} (2005) 023003.
	
	\bibitem {ref10}
	J.W. Maluf,  Ann. Phys. {\bf 525} 339 (2013) 
	\bibitem {ref11}
	S Capozziello, S Nojiri, S D Odintsov and A Troisi, {\it Phys. Lett. B} {\bf 639} 135 (2006)
	
	\bibitem {ref12}
	T. Harko et al., Phys.Rev. D {\bf 84} 024020 (2011)  
	
	\bibitem{ref13}
	M.D. Laurentis, M. Paolella, S. Capozziello, Phys. Rev. D {\bf 91} 083531 (2015)
	
	\bibitem{ref14} J. B. Jimenez et al., Phys. Rev. D {\bf 101} 103507 (2020) 	 
	
	\bibitem {ref15}
	H. Weyl, Sitzungsber. Preuss. Akad. Wiss, Berlin {\bf 465} (1918)
	
	\bibitem {ref16}
	G. Lyra,  Mathematische Zeitschrift {\bf 54} 52 (1951)
	
	\bibitem {ref17}
	C. Brans, R.H. Dicke, Phys. Rev.  {\bf 124} 925 (1961)
	
	\bibitem {ref18}
	H. Yilmaz, Phys. Rev. {\bf 111} 1417 (1958)
	
	\bibitem{Nojiri/2011} S. Nojiri, S. D. Odintsov, Phys. Rep. {\bf 505} (2011) 59.
	
	\bibitem{Nojiri/2017} S. Nojiri, S. D. Odintsov, V. K. Oikonomou, Phys. Rep. {\bf 692} (2017) 1. 
	
	\bibitem{Nojiri/2003prd} S. Nojiri, S. D. Odintsov, Phys. Rev. D {\bf 68} (2003) 123512. 
	
	\bibitem {ref19}
	B. Ratra and P. J. E. Peebles, Phys. Rev. D {\bf 37}, 3406 (1988).
	
	\bibitem {ref20}
	E. J. Copeland, M. Sami, S. Tsujikawa, Int. J. Mod. Phys. D 15, 1753-1936 (2006)
	
	\bibitem {ref21}
	I. Zlatev, L.M. Wang, P.J. Steinhardt, Phys. Rev. Lett. {\bf 82} 896 (1999) 
	
	\bibitem {ref22}
	P.J. Steinhardt, L.M. Wang, I. Zlatev, Phys. Rev. D {\bf 59}, 123504 (1999)
	
	\bibitem {ref23}
	V. B. Johri, Phys. Rev. D {\bf 63}, 103504 (2001).
	
	\bibitem {ref24}
	C. Armendariz-Picon, V.F. Mukhanov, P.J. Steinhardt, Phys. Rev. Lett. {\bf 85}  4438 (2000)
	
	\bibitem {ref25}
	L. Amendola, D. Tocchini-Valentini, Phys. Rev. D {\bf 64} 043509 (2001) 
	
	\bibitem {ref26}
	M. Gasperini, F. Piazza, G. Veneziano, Phys. Rev. D {\bf 65} 023508 (2002) 
	
	\bibitem {ref27}
	L.A. Boyle, R.R. Caldwell, M. Kamionkowski, Phys. Lett. B {\bf 545} 17 (2002)
	
	\bibitem {ref28} 
	A. Jawad, A. Majeed, Astrophy. Space Sc. {\bf 356} 375 (2015)
	
	\bibitem {ref29}
	T. Chiba, T. Okabe, M. Yamaguchi, Phys. Rev. D {\bf 62} 023511 (2000) 
	
	\bibitem {ref30}
	S.H. Mohseni, M. Alimohammadi, Phys. Rev. D {\bf 74.4} 043506 (2006)
	
	\bibitem {ref30a}
	Planck Collaboration, P. A. R. Ade et al., Astron. Astrophys. 594 A 20 (2016)	 
	
	\bibitem {ref31} A. Kamenshchik et al., Phys. Lett. B {\bf 511} 265 (2001)
	
	\bibitem {ref32}
	S. Perlmutter et al., Nature  {\bf 391} 51 (1998)
	
	\bibitem {ref33} C.L. Bennett et al., Astrophys. J. Suppl. 148, 1 (2003)
	
	\bibitem {ref34} G. Hinshaw et al., Astrophys. J. Suppl. 148, 135 (2003)
	
	\bibitem{ref34a} C. Krishnan et al., {\it Phys. Rev. D} {\bf 105.6} (2022) 063514.
	
	\bibitem{ref34b} R. McConville, E.O. Colgain,  arxiv 2304.02718.	
	
	\bibitem{ref35} 
	A.K. Yadav et al., Phys. Dark Univ. {\bf 31} 100738 (2021)
	
	\bibitem {ref36} V.K. Bhardwaj, Mod. Phys. Lett. A {\bf 33} 1850234 (2018)
	
	\bibitem {ref37}
	G.K. Goswami et al., Mod. Phys. Lett. A {\bf 35} 2050086 (2020)
	
	\bibitem{ref38} M. Demianski, et al., Phys. Rev. D {\bf 46} 1391 (1992)
	
	\bibitem{ref39} De, Avik, et al., Euro. Phys. J. C {\bf 82} 72 (2022)
	
	\bibitem{Nojiri/2022NPB} S. Nojiri, S. D. Odintsov, V.K. Oikonomou, A. Constantini, Nucl. Phys. B {\bf 985} 116011 (2022)
	
	\bibitem{ref40} J.K. Singh, R. Nagpal, Euro. Phys. J. C {\bf 80} 295 (2020)
	
	\bibitem {ref41} S Ghaffari et al., Eur. Phys. J. C {\bf 78} 706 (2018)
	
	\bibitem {ref42} O. Akarsu, S. Kumar, S. Sharma, L. Tedesco Phys. Rev. D 100 023532 (2019)
	
	\bibitem {ref43} V.K. Bhardwaj, et al., Chin. J. Phys. {\bf 80} 261 (2022) 
	
	\bibitem {ref44}
	M. Chevallier, D. Polarski, Int. J. Mod. Phys. D {\bf 10} 213 (2001)
	
	\bibitem {ref45}
	E.V. Linder, Phys. Rev. Lett. {\bf 90} 091301 (2003)
	
	\bibitem{ref46} D. Foreman-Mackey et al., Publ. Astron. Soc. Pac. {\bf 125}  306 (2013)
	
	\bibitem{ref47} G. Chen, S. Kumar, B. Ratra, Astrophys. J. 835, 86 (2017)
	
	\bibitem{ref48} G.S. Sharov, V.O. Vasiliev, Math. Model. Geom. 6, 1 (2018)
	
	\bibitem{ref49} D.M. Scolnic et al., Astrophys. J. 859, 101(2018.
	
	\bibitem{ref50} C. Blake et al., Mon. Not. Roy. Astron. Soc. 418, 1707 (2011)
	
	\bibitem{ref51} W. J. Percival et al., Mon. Not. Roy. Astron. Soc. 401, 2148 (2010)
	
	\bibitem{ref52} F. Beutler et al., Mon. Not. Roy. Astron. Soc. 416, 3017 (2011)
	
	\bibitem{ref53} R. Giostri et al., J. Cosmol. Astropart. Phys. {\bf 2012} 027 (2012)
	
	\bibitem{ref54} G. Chen, B. Ratra, Publ. Astron. Soc. Pac. {\bf 123} 1127 (2011)
	
	\bibitem{ref55} E. Aubourg et al., Phys. Rev. D {\bf 92} 123516  (2015)
	
	\bibitem{ref56} G. Hinshaw et al., Astrophys. J. Suppl. Ser. {\bf 208} 25 (2013)
	
	\bibitem{ref56a} A. Banerjee et al., {\it Phys. Rev. D} {\bf 103} (2021) L081305.
	
	\bibitem{ref56b} Bum-Hoon Lee et al., {\it J. Cosmo. Astropart. Phys.} {\bf 2022.04} (2022) 004.
	
	\bibitem{ref56c} P.K. Aluri et al., {\it Class. Quant. Grav.} {\bf 40.9} (2023) 094001.
	
	\bibitem {ref57}
	R.R. Caldwell, Phys. Lett. B {\bf 545} 23 (2002)
	
	\bibitem {ref58}
	M.R. Setare, E.N. Saridakis, Int. J. Mod. Phys. D {\bf 18} 549 (2009)
	
	\bibitem {ref59}
	V.M. Zhuravlev, D.A. Kornilov, E P. Savelova, Gen. Rel. Grav. {\bf 36} 1719 (2004)
	
	\bibitem {ref61}
	H. E. Bond {\it et al.}, Astrophys. J. Lett. {\bf 765}  L12 (2013)
	
	\bibitem {ref62}
	S. Masi et al., Prog. Part. Nucl. Phys. {\bf  48} 243 (2002)
	
	\bibitem {ref63}
	A. Renzini, A. Bragaglia, F.R.  Ferraro, Astrophys. J.,{\bf  465} L23  (1996)
	
	\bibitem {ref64} B.M. Bentabol, J.M. Bentabol, J. Cepa, J. Cosmol. Astropart. Phys. 02, 015 (2013)
	
	\bibitem {ref67}
	R. Nair et al., J. Cosmol. Astropart. Phys. {\bf 01} 018 (2012)
	
	\bibitem {ref68}
	O. Farooq, B. Ratra, Astrophys. J. Lett. {\bf 766}  L7 (2013)
	
	\bibitem {ref69}
	J. Magana et al., J. Cosmol. Astropart. Phys. {\bf 2014(10)} 017 (2014)
	
	\bibitem {ref70}
	A.A. Mamon, K. Bamba, S. Das, Eur. Phys. J. C {\bf 77} 29 (2017)
	
	\bibitem {ref71}
	S. Capozziello, et al., Mon. Not. R. Astron. Soc. {\bf 509} 5399 (2022) 
	
	\bibitem{ref72}
	H. Yu, B. Ratra, F.-Y. Wang, Astrophys. J. {\bf 856} 3 (2018) 
	
	\bibitem {ref73}	
	L. Xu, H. Liu,  Mod. Phys. Lett. A {\bf 23} 1939 (2008)
	
	\bibitem {ref74} 
	A.K. Yadav et al., New Astronomy {\bf 78} 101382 (2020)
	%
	\bibitem {ref75} 
	M. Visser, Class. Quantum Grav. {\bf 21} 2603 (2004)
	
	\bibitem {ref76} 
	L.K. Sharma, A.K. Yadav, B.K. Singh, New Astronomy {\bf 79} 101396 (2020)
	
	\bibitem{ref77}
	R. Nagpal, et al., Ann. Phys. {\bf 405} 234 (2019)
	
	\bibitem{ref78}
	M. Shahalam et al., Mon. Not.  Roy Astron Soc. {\bf 448} 2948 (2015)
	
	\bibitem{ref79}
	R.D. Blandford et al., ASP Conf. Ser. {\bf 339} 27 (2004)  
	
	\bibitem{ref80}
	V. Sahni et al., J. Exp. Theor. Phys. {\bf 77} 201 (2003) 
	
	\bibitem{ref81}
	U. Alam et al., Mon. Not. R. Astron. Soc. {\bf 344} 1057 (2003) 
	
	\bibitem{ref82} J. Santos, J.S. Alcaniz, Phys. Lett. B 619, 11 (2005)
	
	\bibitem{ref83} J. Santos, et al., Phys. Rev. D {\bf 76} 083513 (2007)
	
	\bibitem{ref84} S.M. Carroll, Addison Wesley, Boston, (2004)
	
	\bibitem {ref85}
	S. Capozziello, S. Nojiri and S.D. Odintsov, Phys. Lett. B 781, 99-106 (2018)
	
	\bibitem {ref86}
	S. E. Perez Bergliaffa, Phys. Lett. B 642, 311 (2006)
	
	\bibitem{ref87}
	V. K Bhardwaj, M. K. Rana, A. K Yadav,  Astrophys. Space Sci. {\bf 364} 1 (2019)
	
	\bibitem {ref88} A.K. Yadav, V.K. Bhardwaj, Res. Astron. Astrophys. {\bf 18} 64 (2018) 
	
\end{thebibliography}
\end{document}